\documentclass{PoS}

\title{
An implementation of hybrid parallel CUDA code for the hyperonic nuclear forces
}

\ShortTitle{
An implementation of hybrid parallel CUDA code for the hyperonic nuclear forces
}

\author{\speaker{
Hidekatsu Nemura
}
\\
        Center for Computational Sciences,
        University of Tsukuba, Tsukuba, Ibaraki, 305-8577, Japan\\
        E-mail: \email{
                       nemura.hidekatsu.gb@u.tsukuba.ac.jp
}}

\author{
        for HAL QCD %
	Collaboration\\
	\includegraphics[width=0.20\textwidth]{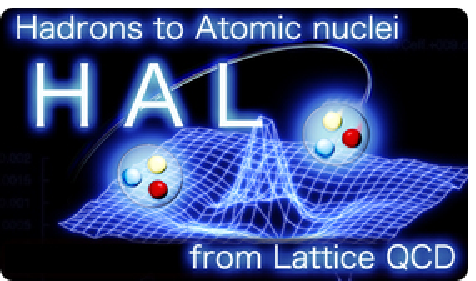}
}

\abstract{
We present our recent effort to develop a GPGPU program to calculate 
52 channels of the 
Nambu-Bethe-Salpeter (NBS) wave functions
in order to study the baryon interactions, 
from nucleon-nucleon to $\Xi-\Xi$, 
from lattice QCD. 
We adopt CUDA programming to perform the multi-GPU execution 
on a hybrid parallel programming with MPI and OpenMP. 
Effective baryon block algorithm is briefly outlined, 
which calculates efficaciously 
a large number of NBS wave functions at the same time, 
and 
three CUDA kernel programs are implemented to materialize the 
effective baryon block algorithm using GPUs 
on 
the single-program multiple-data (SPMD) programming model. 
In order to parallelize multiple GPUs, 
we take both two approaches 
by dividing the time dimension 
and by dividing the spatial dimensions. 
Performances are measured using HA-PACS supercomputer 
in University of Tsukuba, which includes NVIDIA M2090 and 
NVIDIA K20X GPUs. 
Strong scaling and weak scaling measured by using both 
M2090 and K20X GPUs are presented. 
We find 
distinct difference between the M2090 and the K20X 
in the sustained performance measurement of particular kernel executions
which utilize the cudaStream objects. 
}

\FullConference{The 33rd International Symposium on Lattice Field Theory\\
		14 -18 July 2015\\
		Kobe International Conference Center, Kobe, Japan*}

\usepackage{amssymb}
\usepackage{bm}
\usepackage{ascmac}

\newcommand{\Lower}[1]{\smash{\lower 1.5ex \hbox{#1}}}

\begin{document}

\section{Introduction\label{INTRODUCTION}}

Thanks to both 
elevating 
computer performance and
various inventions of numerical algorithms,
the lattice QCD approach to nuclear physics 
is being developed as a first-principle calculation. 
Not only
two-body systems~\cite{Fukugita:1994ve,Beane:2006mx,Muroya:2004fz,BeaneYN2007}
but also nuclear few-body systems~\cite{Yamazaki:2009ua,Beane:2012vq} 
are the playground for the present-day lattice QCD simulations. 
In addition, 
a new approach to the $NN$ interaction from the lattice 
QCD has been proposed\cite{Ishii:2006ec,Aoki:2009ji}. 
In this approach, the nucleon-nucleon ($NN$) potential can be obtained 
from the lattice QCD by measuring the Nambu-Bethe-Salpeter (NBS) wave function and the 
observables such as the phase shifts and the binding energies are calculated through
the resultant potential\cite{Aoki:2012tk}. 
This approach has been further extended 
and applied to various hadronic systems. 
See Ref.~\cite{Sasaki:2015ifa} and references therein for the 
state-of-the-art 
outcomes. 
Furthermore, 
a large scale lattice QCD calculation is now in 
progress~\cite{DoiIshiiSasaki2015LAT} 
to study the baryon interactions 
from $NN$ to $\Xi\Xi$ by measuring the NBS wave functions 
for 52 channels.

The purpose of this paper is to present 
our recent effort to develop a hybrid parallel 
GPGPU program for multiple devices 
to perform the calculation of the baryon interactions. 
This report is organized as follows:
Section~\ref{FORMULATION}
briefly outlines the effective block algorithm
to calculate the NBS wave functions. 
Section~\ref{HAPACS}
shows 
the machine and programming softwares used 
in this work. 
Section~\ref{IMPLECUDA} is devoted to 
present 
the hybrid parallel program for 
multi-GPU calculation. 
In Sec.~\ref{SEC:RESULTS} we show the performances of the 
calculation with GPUs for the 
52 channels of the NBS wave functions. 
Sec.~\ref{SUMMARY} summarizes the report.


\section{Formulation\label{FORMULATION}}

In order to study the baryon interactions, 
the primary quantity we 
compute with 
lattice QCD is the four-point correlation function defined by
\begin{equation}
 { F}_{\alpha_{1}\alpha_{2},\alpha_{3}\alpha_{4}}
     ^{\langle B_1B_2\overline{B_3B_4}\rangle}(\vec{r},t-t_0) = 
 \sum_{\vec{X}}
 \left\langle 0
  \left|
   B_{1,{\alpha_1}}(\vec{X}+\vec{r},t)
   B_{2,{\alpha_2}}(\vec{X},t)
   \overline{{\cal J}_{B_{3,\alpha_{3}} B_{4,\alpha_{4}}}
                     (t_0)}
  \right| 0 
 \right\rangle,
\end{equation}
where the summation over $\vec{X}$ selects  states with 
zero total momentum. 
The $B_{1,\alpha_1}(x)$ and $B_{2,\alpha_2}(y)$ denote the 
interpolating fields of the baryons such as 
%
\begin{eqnarray}
 &&
 \begin{array}{lll}
  p = \varepsilon_{abc} \left(
			 u_a C\gamma_5 d_b
			\right) u_c,
  &
  n = - \varepsilon_{abc} \left(
			   u_a C\gamma_5 d_b
			  \right) d_c,
  \quad
  &
  \Lambda = {1\over \sqrt{6}} \left( X_u + X_d - 2 X_s \right), 
  \\
  \Sigma^{+} = - \varepsilon_{abc} \left(
				    u_a C\gamma_5 s_b
				   \right) u_c,
  \quad
  &
  \Sigma^{0} = {1\over\sqrt{2}} \left( X_u - X_d \right), 
  &
  \Sigma^{-} = - \varepsilon_{abc} \left(
				    d_a C\gamma_5 s_b
				   \right) d_c,
  \\
  \Xi^{0} = \varepsilon_{abc} \left(
                               u_a C\gamma_5 s_b
                              \right) s_{c},
  &
  \Xi^{-} = - \varepsilon_{abc} \left(
                                 d_a C\gamma_5 s_b
                                \right) s_{c},
  &
 \end{array}
 \label{BaryonOperatorsOctet}
\end{eqnarray}
where
\begin{equation}
 \begin{array}{l}
  X_u = \varepsilon_{abc} \left( d_a C\gamma_5 s_b \right) u_c, 
  \quad 
  X_d = \varepsilon_{abc} \left( s_a C\gamma_5 u_b \right) d_c,
  \quad 
  X_s = \varepsilon_{abc} \left( u_a C\gamma_5 d_b \right) s_c.
 \end{array}
 \label{BaryonOperatorsXuXdXs}
\end{equation}
%
For simplicity, we have suppressed the explicit spinor indices and 
spatial coordinates in Eqs.~(\ref{BaryonOperatorsOctet}) and 
(\ref{BaryonOperatorsXuXdXs}); 
%
$\overline{{\cal J}_{B_{3,\alpha_{3}}B_{4,\alpha_{4}}}
                   (t_0)}$ 
is a source operator which create $B_{3,\alpha_{3}}B_{4,\alpha_{4}}$ states 
at $t=t_{0}$.
Hereafter, explicit time dependences are suppressed. 
In order to quantify the four-point correlation function 
${F}^{\langle B_1B_2\overline{B_3B_4}\rangle}
    _{{\alpha_1}{\alpha_2},{\alpha_3}{\alpha_4}}(\vec{r})$,
we first consider the Wick's contraction 
together with defining the baryon blocks 
$[B_{1,\alpha_{1}}^{(0)}](\vec{x};\xi_{P_{1}}^{\prime}, \xi_{P_{2}}^{\prime}, \xi_{P_{3}}^{\prime})$ and 
$[B_{2,\alpha_{2}}^{(0)}](\vec{y};\xi_{P_{4}}^{\prime}, \xi_{P_{5}}^{\prime}, \xi_{P_{6}}^{\prime})$, 
%
\begin{equation}
 \begin{array}{rcl}
 { F}
     ^{\langle B_1B_2\overline{B_3B_4}\rangle}
     _{{\alpha_1}{\alpha_2},{\alpha_3}{\alpha_4}}(\vec{r}%
                                                          )
 &=& 
  \sum_{\vec{X}} 
   \sum_{P} \sigma_{P}
   ~
   [B_{1,\alpha_1}^{(0)}](\vec{X}+\vec{r};
   ~
   \xi^{\prime}_{P_{1}}, \xi^{\prime}_{P_{2}}, \xi^{\prime}_{P_{3}}
                         ) 
   ~
   [B_{2,\alpha_2}^{(0)}](\vec{X};
   ~
   \xi^{\prime}_{P_{4}}, \xi^{\prime}_{P_{5}}, \xi^{\prime}_{P_{6}}
                         )
 \\
 &&
 \qquad 
 \times 
 \varepsilon_{c_{{1}}^{\prime} c_{{2}}^{\prime} c_{{3}}^{\prime}}
 \varepsilon_{c_{{4}}^{\prime} c_{{5}}^{\prime} c_{{6}}^{\prime}}
 (C\gamma_5)_{\alpha_{1}^{\prime}\alpha_{2}^{\prime}}
 (C\gamma_5)_{\alpha_{4}^{\prime}\alpha_{5}^{\prime}}
 \delta_{\alpha_{3}^{\prime}\alpha_{3}}
 \delta_{\alpha_{6}^{\prime}\alpha_{4}},
 \end{array}
 \label{NaiveBaryonBlocks}
\end{equation}
%
with 
\begin{equation}
 \begin{array}{l}
  ~
  [B_{1,\alpha_{1}}^{(0)}](\vec{x};~
   \xi_{P_{1}}^{\prime}, \xi_{P_{2}}^{\prime}, \xi_{P_{3}}^{\prime}
                        ) 
  =
  \left\langle 
   B_{1,\alpha_{1}}(\vec{x})
   ~
   \bar{q}_{B_{1},3}^{\prime}(\xi_{P_{3}}^{\prime})
   \bar{q}_{B_{1},2}^{\prime}(\xi_{P_{2}}^{\prime})
   \bar{q}_{B_{1},1}^{\prime}(\xi_{P_{1}}^{\prime})
 \right\rangle,
 \quad \mbox{and} 
 \\
 ~
  [B_{2,\alpha_{2}}^{(0)}](\vec{y};~
   \xi_{P_{4}}^{\prime}, \xi_{P_{5}}^{\prime}, \xi_{P_{6}}^{\prime}
                        ) 
  =
  \left\langle 
   B_{2,\alpha_{2}}(\vec{y})
   ~
   \bar{q}_{B_{2},6}^{\prime}(\xi_{P_{6}}^{\prime})
   \bar{q}_{B_{2},5}^{\prime}(\xi_{P_{5}}^{\prime})
   \bar{q}_{B_{2},4}^{\prime}(\xi_{P_{4}}^{\prime})
   \right\rangle,
 \end{array}
\end{equation}
%
where $\sigma_{P}$ and 
$\{\xi_{P_{1}}^{\prime},\cdots,\xi_{P_{6}}^{\prime}\}$ 
are the sign factor and the set of permutated 
color-spin-space coordinates 
for each permutation $P$, 
respectively. 
Both 3-tuple sets of the quark fields 
$\{\bar{q}_{B_{1},1}^{\prime},\bar{q}_{B_{1},2}^{\prime},
   \bar{q}_{B_{1},3}^{\prime}\}$ and 
$\{\bar{q}_{B_{2},4}^{\prime},\bar{q}_{B_{2},5}^{\prime},
   \bar{q}_{B_{2},6}^{\prime}\}$ 
are ordered 
properly 
so as to 
correspond to 
the $B_{1}$ and $B_{2}$ states. 
Taking the expression in Eq.~(\ref{NaiveBaryonBlocks}), 
the number of the iterations to obtain a $
 { F}
     ^{\langle B_1B_2\overline{B_3B_4}\rangle}
     _{{\alpha_1}{\alpha_2},{\alpha_3}{\alpha_4}}(\vec{r}%
                                                          )$ 
reduces to 
$(N_{c}!N_{\alpha})^{B}\times 
N_{u}!N_{d}!N_{s}!\times 
2^{N_{\Lambda}+N_{\Sigma^{0}}-B}$, 
where $N_{c}=3, N_{\alpha}=4$ and 
$N_{\Lambda}$, $N_{\Sigma^{0}}$, 
$N_{u},N_{d},N_{s}$ and $B$ are the numbers of 
$\Lambda,\Sigma^{0}$, up-quark, down-quark, strange-quark and the baryons 
(i.e., always $B=2$ in the present study), respectively. 
In Ref.~\cite{Nemura0806.1094}, only the limited spatial points 
were 
evaluated because of the 
computational 
cost $O(L^{6})$ 
in the primitive numerical approach. 
After that 
we employed the Fast-Fourier-Transform (FFT) and the effective baryon blocks 
to improve the numerical performance to $O(L^3\log L^3)$~\cite{Nemura:2009kc};
%
\begin{equation}
 { F}
     ^{\langle B_1B_2\overline{B_3B_4}\rangle}
     _{{\alpha_1}{\alpha_2},{\alpha_3}{\alpha_4}}\!(\vec{r}%
                                                          )
 \!=\!\!
  \sum_{P} \!\!\sigma_{P}
   \!\sum_{\vec{X}} 
    \!\!\left(\!
         [B_{1,\alpha_1}^{(P)}](\vec{X}\!\!+\!\!\vec{r}) \!\!\times\!\!
         [B_{2,\alpha_2}^{(P)}](\vec{X})
         \!\right)\!\!_{\alpha_3\alpha_4%
                                   }
  \!\!\!=\!
  {1\over L^3}
   \!\!\sum_{\vec{q}}
    \!\!\left(\!\!
     \sum_{P} \!\!\sigma_{P}
      \!\!\left(\!\!
           [\widetilde{B_{1,\alpha_1}^{(P)}}]( \vec{q}) \!\!\!\times\!\!\! 
           [\widetilde{B_{2,\alpha_2}^{(P)}}](\!\!-\vec{q})
      \!\!\!\right)\!\!_{\alpha_3\alpha_4}
     \!\!\!\right)\!\!
     {\rm e}^{i\vec{q}\cdot\vec{r}}.
  \label{B1B2.B1B2B4B3.FFT}
\end{equation}
%
See Ref.~\cite{Nemura:2015yha} how to extract the 
effective baryon blocks. 
For example, 
the specific form of the four-point correlation function 
${F}^{\langle p \Lambda\overline{pX_{u}}\rangle}
    _{{\alpha_1}{\alpha_2},{\alpha_3}{\alpha_4}}(\vec{r})$ 
of the $\langle p \Lambda\overline{pX_{u}}\rangle$ channel 
is given by, 
%
\begin{eqnarray}
 \!\!\!\!{F}^{\langle p \Lambda\overline{pX_{u}}\rangle}
    _{{\alpha_1}{\alpha_2},{\alpha_3}{\alpha_4}}(\vec{r})
 \!\!\!&=&\!\!\!
  {1\over L^3}
   \sum_{\vec{q}}
    \left(
    [\widetilde{      p}_{\alpha_1\alpha_3}^{(1)}]( \vec{q}) 
    [\widetilde{\Lambda}_{\alpha_2\alpha_4}^{(1)}](-\vec{q})
    -
    [\widetilde{      p}_{\alpha_1\alpha_4}^{(2)}]_{c_{3}^{\prime}c_{6}^{\prime}}( \vec{q}) 
    [\widetilde{\Lambda}_{\alpha_2\alpha_3}^{(2)}]_{c_{3}^{\prime}c_{6}^{\prime}}(-\vec{q})
    \right.
    \nonumber
    \\
    && \!\!\!\!\!\!
    \left.
    -
    [\widetilde{      p}_{\alpha_1\alpha_3}^{(3)}]_{c_{2}^{\prime}\alpha_{2}^{\prime}c_{4}^{\prime}\alpha_{4}^{\prime}}( \vec{q}) 
    [\widetilde{\Lambda}_{\alpha_2\alpha_4}^{(3)}]_{c_{2}^{\prime}\alpha_{2}^{\prime}c_{4}^{\prime}\alpha_{4}^{\prime}}(-\vec{q})
    \right.
    \left.
    +
    [\widetilde{      p}_{\alpha_1\alpha_4}^{(4)}]_{c_{1}^{\prime}\alpha_{1}^{\prime}c_{5}^{\prime}\alpha_{5}^{\prime}}( \vec{q}) 
    [\widetilde{\Lambda}_{\alpha_2\alpha_3}^{(4)}]_{c_{1}^{\prime}\alpha_{1}^{\prime}c_{5}^{\prime}\alpha_{5}^{\prime}}(-\vec{q})
    \right.
    \nonumber
    \\
    && \!\!\!\!\!\!
    \left.
    +
    [\widetilde{      p}_{\alpha_1\alpha_3\alpha_4}^{(5)}]_{c_{1}^{\prime}\alpha_{1}^{\prime}c_{6}^{\prime}}( \vec{q}) 
    [\widetilde{\Lambda}_{\alpha_2                }^{(5)}]_{c_{1}^{\prime}\alpha_{1}^{\prime}c_{6}^{\prime}}(-\vec{q})
    \right.
    \left.
    -
    [\widetilde{      p}_{\alpha_1\alpha_3\alpha_4}^{(6)}]_{c_{3}^{\prime}c_{5}^{\prime}\alpha_{5}^{\prime}}( \vec{q}) 
    [\widetilde{\Lambda}_{\alpha_2                }^{(6)}]_{c_{3}^{\prime}c_{5}^{\prime}\alpha_{5}^{\prime}}(-\vec{q})
    \right)
    {\rm e}^{i\vec{q}\cdot\vec{r}}.
  \label{pL.pLXup.FFT}
\end{eqnarray}
%
By employing the effective block algorithm, 
the number of iterations to evaluate the r.h.s. of Eq.~(\ref{pL.pLXup.FFT})
except the momentum space degrees of freedom becomes 
$1+N_{c}^{2}+N_{c}^{2}N_{\alpha}^{2}+N_{c}^{2}N_{\alpha}^{2}+
N_{c}^{2}N_{\alpha}+N_{c}^{2}N_{\alpha}=370$, 
which is significantly smaller than the number 
$(N_{c}!N_{\alpha})^{B}\times 
N_{u}!N_{d}!N_{s}!\times 
2^{N_{\Lambda}+N_{\Sigma^{0}}-B} = 3456$ 
seen in Eq.~(\ref{NaiveBaryonBlocks}). 
The manipulation on the expression of Eq.~(\ref{B1B2.B1B2B4B3.FFT}) 
in terms of the effective blocks 
$[B_{1,\alpha_{1}}^{(P)}]$ and $[B_{2,\alpha_{2}}^{(P)}]$
can be automatically done once the set of the interpolating fields 
(i.e., the quantum numbers) 
of both sink and source parts is given~\cite{Nemura:2015yha}.

\section{Machine and programming softwares\label{HAPACS}}

The present implementation is performed to utilize HA-PACS supercomputer
in University of Tsukuba, 
which 
includes 
base cluster part and 
tightly coupled accelerators (TCA) part. 
The base cluster part 
consists of 
268 %
nodes, each of which comprises 
two Intel E5-2670 CPUs as well as %
four NVIDIA M2090 GPUs %
connected by PCI-express, and started for common use in 2012. 
The TCA part 
involving 64 %
nodes 
was added 
to the HA-PACS 
in 2013, 
each of which comprises 
two Intel E5-2680v2 CPUs %
and 
four NVIDIA K20X GPUs.%
~\footnote{
The TCA system is developed 
to implement 
a proprietary interconnect especially for accelerators, in
order to shorten the communication latency among accelerators
over different nodes\cite{TCA2013IEEE}.
}
%
Table~\ref{SpecsHAPACS} summarizes %
properties of these GPUs. 
\begin{table}[t]
 \begin{minipage}{\textwidth}
  \begin{center}
  \footnotesize
\begin{tabular}{lll}
 \hline
                          & Base cluster part   & TCA part           \\
 \hline
 Name                     & Tesla M2090         & Tesla K20Xm        \\
 Peak performance (GFlops, DP)
                          & 665                 & 1310               \\
 Compute capability       & 2.0                 & 3.5                \\
 Global memory (GiB)      & 5.25                & 5.62               \\
 ECC                      & Enabled             & Enabled        \\
 Clock rate (GHz)         & 1.30                & 0.732              \\
 Memory bus width (bit)   & 384                 & 384                \\
 Memory clock rate (GHz)  & 1.85                & 2.60               \\
 Constant memory (KiB)    & 64                  & 64                 \\
 Shared memory per block 
                    (KiB) & 48                  & 48                 \\
 32-bit registers available per block
                          & 32768               & 65536               \\
 Threads in warp          & 32                  & 32                  \\
 \hline
\end{tabular}
   \caption{Several outputs from {\it cudaDeviceProp} and 
            the peak performance values of double precision (DP) in 
            GFlops obtained from HA-PACS. 
           }
   \label{SpecsHAPACS}
  \end{center}
 \end{minipage}
\end{table}
%
For programming softwares on HA-PACS in this report we employed 
Intel C++ Compiler Version 14.0.4.211, 
Intel MPI Library 4.1 for Linux,
and 
NVIDIA Cuda compiler driver version 6.5.14. 

\section{Implementation of hybrid parallel CUDA code for multiple GPUs system
\label{IMPLECUDA}}

In Ref.~\cite{Nemura:2015yha},
we developed a hybrid parallel C++ program to 
calculate 
the 52 channels of 
the four-point correlation functions by using both MPI and OpenMP. 
The program works on either Bridge++ or CPS++, 
where both modified versions are 
employed. 
For Bridge++, 
feasibility of two frameworks, OpenCL and OpenACC, 
to utilize the GPU 
is discussed~\cite{Motoki:2015LAT}.
In this work, 
we adopt NVIDIA's CUDA programming for the first testbed implementation, 
because 
the target machine is HA-PACS comprising NVIDIA's Fermi and Kepler 
generation GPUs so that 
a better performance is expected with developing the 
CUDA programming than others. 
We also, in this work, 
aim to materialize the 
multi-GPU execution 
on the hybrid parallel programming with 
MPI and OpenMP 
by the single-program multiple-data (SPMD) programming model. 
In order to utilize multiple GPUs from a single program, 
we assign one MPI process to each GPU. 
Basic arithmetic of double-precision complex-numbers is implemented 
by hand with utilizing shared memory. 
Constant memory is employed to store 
runtime parameters and constant parameters. 
In the following, 
we describe three CUDA kernels implemented to 
calculate the 52 channels of the NBS wave functions.

\noindent
{\bf (i)}~{$\bm{NormalBaryonBlocks}$}
:~
Assuming that the quark propagators are already solved, 
we first compute the normal baryon blocks on GPUs:
\begin{equation}
  [B_{\alpha}^{(0)}](\vec{r};~
   \xi_{1}^{\prime}, \xi_{2}^{\prime}, \xi_{3}^{\prime}
                        )
  =
  \left\langle 
   B_{\alpha}(\vec{r})
   ~
   \bar{q}_{3}^{\prime}(\xi_{3}^{\prime})
   \bar{q}_{2}^{\prime}(\xi_{2}^{\prime})
   \bar{q}_{1}^{\prime}(\xi_{1}^{\prime})
 \right\rangle,
\quad
\mbox{with}
\quad
B=p,\Sigma^{+},\Xi^{0},X_{u},X_{d},X_{s}, 
\end{equation}
where 
three quark flavors $q_{1}^{\prime},q_{2}^{\prime},q_{3}^{\prime}$ 
are appropriately chosen to create the corresponding $B$ state. 
The other baryon blocks, 
$B=n,\Sigma^{-},\Xi^{-},\Sigma^{0},\Lambda$, 
are obtained from the above according to Eqs.~(\ref{BaryonOperatorsOctet}) 
and (\ref{BaryonOperatorsXuXdXs}) 
with presuming the symmetricity under the 
interchange of up and down quarks in the isospin symmetric limit. 
After the kernel execution, 
the 
FFT is employed 
to obtain the baryon blocks in momentum space. 
No clear benefit nor clear disadvantage is 
observed in performing the FFT whether on host side or on device side; 
the bottleneck is due to the {\it Alltoall} MPI communications 
for the FFT. 
The data of normal baryon blocks are replaced by its in momentum space 
after the FFT.

\noindent
{\bf (ii)}~{$\bm{EffectiveBaryonBlocks}$}
:~
We 
construct the effective baryon blocks 
from the 
normal baryon blocks 
in momentum space, 
\begin{eqnarray}
  \left\{
  [\widetilde{B_{1,\alpha_1}^{(d)}}]_{\tilde{\bm{\xi}}_d}( \vec{q}),
  [\widetilde{B_{2,\alpha_2}^{(d)}}]_{\tilde{\bm{\xi}}_d}(-\vec{q});
  \alpha_{3},\alpha_{4}
  \right\},
  \qquad
  \mbox{with}
  \qquad
  B=p,\Sigma^{+},\Xi^{0},X_{u},X_{d},X_{s}, 
\end{eqnarray}
where $\{\tilde{\bm{\xi}}_d\}$ denotes the indices 
which originate from the quark fields in the source; 
for example, in Eq.~(\ref{pL.pLXup.FFT}), 
it becomes 
$\{\mbox{none}\}$, 
$\{c_{3}^{\prime},c_{6}^{\prime}\}$,
$\{c_{2}^{\prime},\alpha_{2}^{\prime},c_{4}^{\prime},\alpha_{4}^{\prime}\}$,
$\{c_{1}^{\prime},\alpha_{1}^{\prime},c_{5}^{\prime},\alpha_{5}^{\prime}\}$,
$\{c_{1}^{\prime},\alpha_{1}^{\prime},c_{6}^{\prime}\}$,
$\{c_{3}^{\prime},c_{5}^{\prime},\alpha_{5}^{\prime}\}$ 
for the six terms of the four-point correlator of the channel 
$\langle p \Lambda\overline{pX_{u}}\rangle$. 
In order to avoid the warp divergence, 
the diagramatical classification is performed throughout in the CPU code 
and the resultant data is aligned in 
Structure of Arrays (SoA) format which is transferred to the device 
prior to the kernel execution. 
This kernel execution 
has 
less timing performance impact 
though it is indispensable to connect the former part and the next part. 
Therefore we have not paid 
very 
much attention to improve the performance 
of this kernel.

\noindent
{\bf (iii)}~{$\bm{MultiplicationEffectiveBlocks}$}
:~
Performed the kernel executions described in the above, 
we make the product of two effective baryon blocks on 
GPUs, 
%
\begin{eqnarray}
  \left(
       [\widetilde{B_{1,\alpha_1}^{(d)}}]( \vec{q}) \times 
       [\widetilde{B_{2,\alpha_2}^{(d)}}](-\vec{q})
  \right)_{\alpha_{3}\alpha_{4}}
 &=&
  \sum_{\tilde{\bm{\xi}}_d}
  \left(
      ~
      [\widetilde{B_{1,\alpha_1}^{(d)}}]_{\tilde{\bm{\xi}}_d}( \vec{q}) 
      ~
      [\widetilde{B_{2,\alpha_2}^{(d)}}]_{\tilde{\bm{\xi}}_d}(-\vec{q})
  \right)_{\alpha_{3}\alpha_{4}}
.
 \label{ProductOfEffectiveBaryonBlocks}
\end{eqnarray}
%
For entire 52 channels of the NBS wave functions, 
we have to consider all of such summations tabulated in Tables~1-4 in 
Ref.~\cite{Nemura:2015yha}. 
This is one of the time consuming part of the present calculation. 
The symbolical manipulations are 
performed in the CPU code and the resultant SoA formed data is 
transferred to GPU to suppress the warp divergence 
prior to the kernel execution. 
To have good overlapping the kernel executions and 
the data transfer between the host and device, 
especially for the K20X GPU, 
we also utilize the 
{\it cudaStream} objects. 
The NBS wave function is finally obtained by performing the 
inverse FFT. 


\section{Results\label{SEC:RESULTS}}

\begin{figure}[t]
 \begin{minipage}[t]{0.49\textwidth}
  \centering \leavevmode 
  \includegraphics[width=0.99\textwidth]{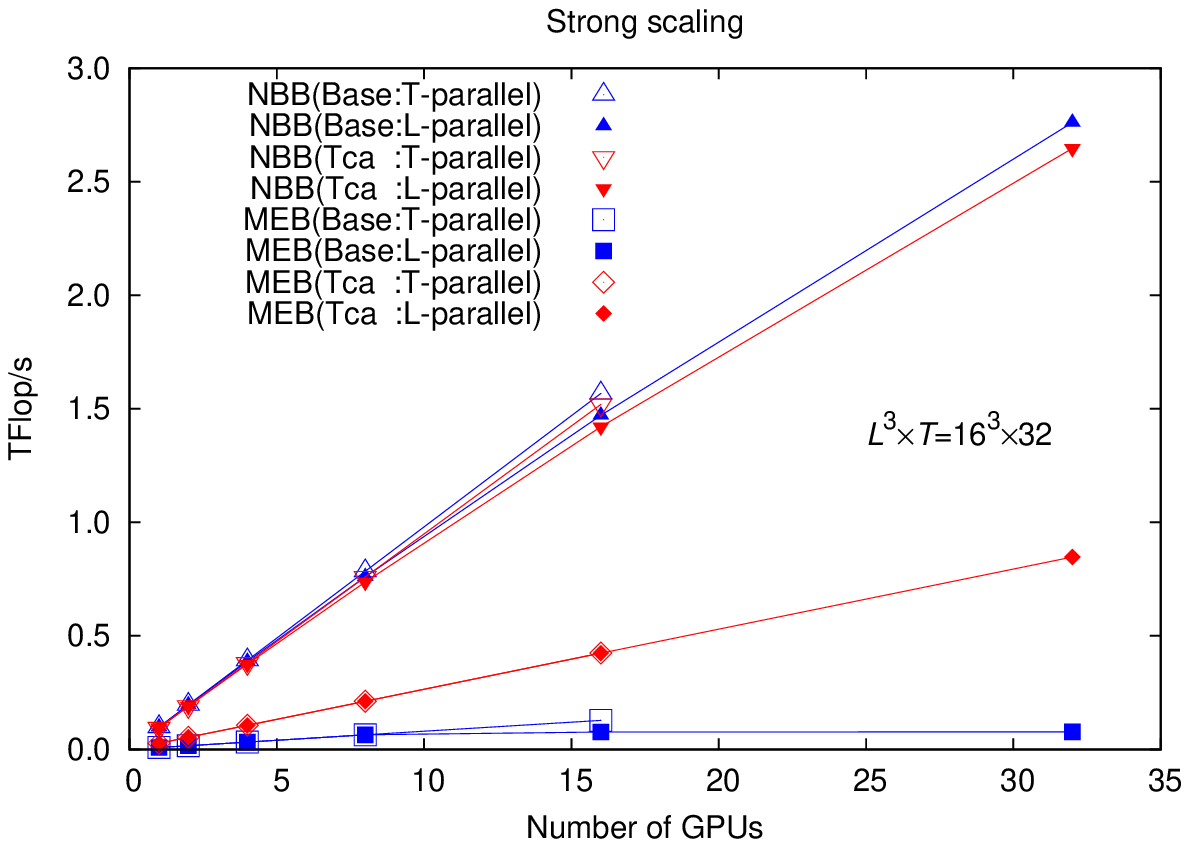}
  \footnotesize
 \caption{Strong scaling of the CUDA kernel executions 
   with the 
   lattice size $L^3\times T=16^3\times 32$. 
 \label{Fig_Strong}}
 \end{minipage}~
 \hfill
 \begin{minipage}[t]{0.49\textwidth}
  \centering \leavevmode 
 \includegraphics[width=0.99\textwidth]{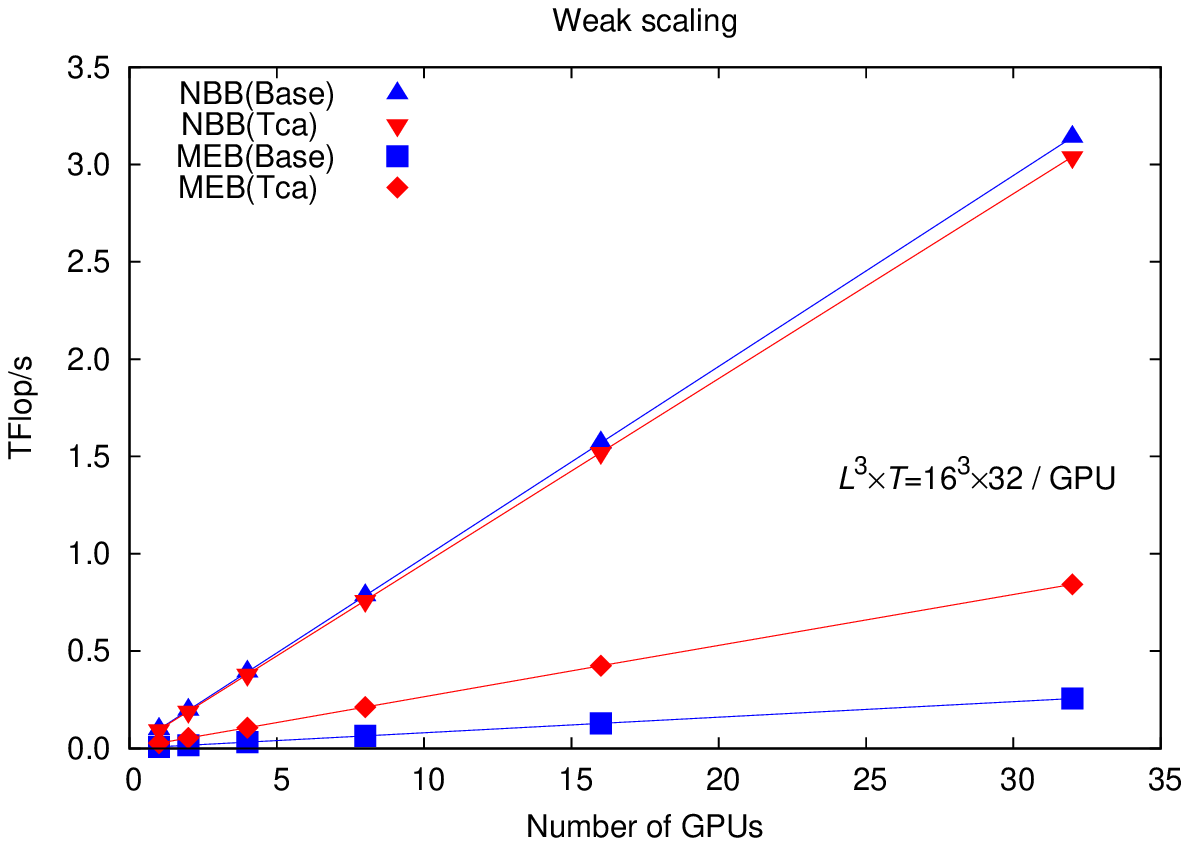}
  \footnotesize
 \caption{Weak scaling of the CUDA kernel executions with the 
   lattice size $L^3\times T=16^3\times 32$ per GPU. 
 \label{Fig_Weak}}
 \end{minipage}
\end{figure}

\begin{table}[b]
 \begin{minipage}{\textwidth}
  \begin{center}
  \footnotesize
\begin{tabular}{lcc}
 \hline
 Kernel                              & Tesla M2090      & Tesla K20Xm \\
 \hline
 {\it NormalBaryonBlocks}            & 98 (98)          & 94 (95)  \\
 {\it EffectiveBaryonBlocks}         & 3.1~\quad~       & 0.55~\quad~ %
                                                                                                              \\
 {\it MultiplicationEffectiveBlocks} & 8.2 (8.0)        & 3.2 (27)    %
                                                                                                              \\
 \hline
\end{tabular}
   \caption{The performance values in GFlops using single GPU 
            for each single kernel execution of 
            {\it NormalBaryonBlocks}, 
            {\it EffectiveBaryonBlocks} or 
            {\it MultiplicationEffectiveBlocks}, 
            measured on the base cluster part (M2090) or 
            on the TCA part (K20X). 
            The calculation is performed 
            in an accuracy of double precision 
            with 
            the lattice size $L^3\times T=16^3\times 32$.
            In parentheses 
            the sustained performance values handled by cudaStream 
            are shown for the kernel executions of 
            {\it NormalBaryonBlocks} and {\it MultiplicationEffectiveBlocks}
            in GFlops. 
   }
   \label{RESULTS_FLOPS}
  \end{center}
 \end{minipage}
\end{table}
Table~\ref{RESULTS_FLOPS} shows the performance values 
of double-precision computation 
in GFlops 
using single GPU for each single kernel execution of 
{\it NormalBaryonBlocks}, 
{\it EffectiveBaryonBlocks} 
or 
{\it MultiplicationEffectiveBlocks} 
with 
$L^3\times T=16^3\times 32$ lattice 
measured on the base cluster part (M2090) 
or 
the TCA part (K20X). 
We also list the sustained GFlops values handled by {\it cudaStream} 
for the kernel executions of 
{\it NormalBaryonBlocks} and {\it MultiplicationEffectiveBlocks} 
in parentheses. 
The 
handling 
of kernel executions 
for {\it MultiplicationEffectiveBlocks} 
by {\it cudaStream} 
lifts up the performance for the K20X more than factor 8 
whereas no 
improvement 
is observed for the M2090. 
This is because the different 
architecture between the M2090 and the K20X; 
the compute capability of M2090 (K20X) is 2.0 (3.5). 
Figure~\ref{Fig_Strong} shows the strong scalings of 
two kernel executions {\it NormalBaryonBlocks} (NBB) and 
{\it MultiplicationEffectiveBlocks} (MEB) 
with total lattice size $L^3\times T=16^3\times 32$ 
measured on the base cluster part (M2090) and the TCA part (K20X). 
In parallelizing across multiple GPUs, we take both two approaches 
by dividing the time 
dimension 
and by dividing the spatial 
dimensions, 
which are indicated by ``T-parallel'' 
and ``L-parallel'' 
in the figure. 
Detailed parameters which specifies the load on each GPU 
are adjusted on each measurement. 
Figure~\ref{Fig_Weak} shows the weak scalings of 
two kernel executions {\it NormalBaryonBlocks} (NBB) and 
{\it MultiplicationEffectiveBlocks} (MEB) 
for (an-)isotropic lattice with size $L^3\times T=16^3\times 32$ 
per GPU 
measured on the base cluster part (M2090) and the TCA part (K20X). 
%

\section{Summary\label{SUMMARY}}

In this paper, we present a recent effort to develop the 
hybrid parallel GPGPU program 
for multiple devices 
that calculates the 52 channels of the NBS wave functions. 
The implementation and the performance measurements are performed by using 
HA-PACS supercomputer in University of Tsukuba, 
which comprises 
the base cluster part including NVIDIA M2090 and 
the TCA part including NVIDIA K20X. 
In order to have better performance by using the GPUs, 
we adopt CUDA programming for the first testbed implementation. 
In performing the FFT, 
no clear benefit nor clear disadvantage is 
observed whether by using CPUs or by using GPUs. 
Three kernel programs are implemented by considering the 
data ordering on the device memory, 
suppression of the warp divergence, 
and making use of 
the shared memory and the constant memory. 
We also employ the cudaStream to perform efficiently the 
kernel executions as well as 
the data transfers between the host and device 
because the data transfers are indispensable 
for the large scale calculation of 52 channels of the NBS wave functions. 
The strong scaling and the weak scaling are measured for the 
kernel executions. 
Distinct difference between the M2090 and the K20X is observed 
in the sustained performance measurement of the particular kernel executions;
handling of kernel executions by using cudaStream is a key to 
make better use of latest GPUs in this approach.

\acknowledgments

The author would 
like to thank  
CP-PACS/JLQCD collaborations %
and 
ILDG/JLDG~\cite{ILDGJLDG} 
for 
allowing us to access the full QCD gauge configurations, 
and developers of Bridge++~\cite{BRIDGEPLUSPLUS},
and the Computational Materials Science Initiative (CMSI). 
Calculations in this paper have been performed 
by using the HA-PACS computer under the 
Interdisciplinary 
Computational Science Program in CCS, University of Tsukuba. 
This research was supported in part 
by Strategic Program for Innovative Research (SPIRE),
the MEXT Grant-in-Aid,  
Scientific Research on Innovative Areas 
(No. 25105505).

\end{document}